\RequirePackage{fix-cm} % Fix LaTeX2e bugs.
\documentclass[a4paper, twoside, reqno, 12pt, dvips]{amsart}

\usepackage{fixltx2e}     % Fix LaTeX2e bugs.

\usepackage{amssymb}
\usepackage{amsmath}
\usepackage{mathrsfs, dsfont}

\usepackage{a4wide}

\usepackage{acronym}

\usepackage{indentfirst}
\usepackage{graphicx}
\usepackage{psfrag}

\usepackage[usenames,dvipsnames]{pstricks}
\usepackage{epsfig}
\usepackage{pst-grad} % For gradients
\usepackage{pst-plot} % For axes

\usepackage{subfigure}

\usepackage{longtable}

\usepackage[english]{babel}
\usepackage[latin1]{inputenc}

\usepackage{color}
\definecolor{oneblue}{rgb}{0,0.0,0.75}
\usepackage[colorlinks,
            urlcolor=oneblue,
            linkcolor=oneblue,
            citecolor=oneblue,
            bookmarksopen=false,
            pagebackref]{hyperref}

\numberwithin{equation}{section}

\newcommand{\R}{\mathbb{R}}
\newcommand{\I}{\mathcal{I}}
\newcommand{\od}[2]{\frac{d#1}{d#2}}
\newcommand{\sign}{\mathop{\mathrm{sign}}}

\begin{document}

\title[Waves generated by an underwater landslide]{Dispersive waves generated by an underwater landslide}

\author[D. Dutykh]{Denys Dutykh$^*$}
\address{LAMA, UMR 5127 CNRS, Universit\'e de Savoie, Campus Scientifique, 73376 Le Bourget-du-Lac Cedex, France}
\email{Denys.Dutykh@univ-savoie.fr}
\urladdr{http://www.lama.univ-savoie.fr/~dutykh/}
\thanks{$^*$ Corresponding author}

\author[D. Mitsotakis]{Dimitrios Mitsotakis}
\address{IMA, University of Minnesota, 114 Lind Hall and 207 Church Street SE, Minneapolis MN 55455, USA}
\email{dmitsot@gmail.com}
\urladdr{http://sites.google.com/site/dmitsot/}

\author[S. Beysel]{Sonya Beysel}
\address{Institute of Computational Technologies, Siberian Branch of the Russian Academy of Sciences, 6 Acad. Lavrentjev Avenue, 630090 Novosibirsk, Russia}
\email{beisels@gmail.com}

\author[N. Shokina]{Nina Shokina}
\address{Institute of Computational Technologies, Siberian Branch of the Russian Academy of Sciences, 6 Acad. Lavrentjev Avenue, 630090 Novosibirsk, Russia}
\email{nina.shokina@googlemail.com}

\begin{abstract}
In this work we study the generation of water waves by an underwater sliding mass. The wave dynamics are assumed to fell into the shallow water regime. However, the characteristic wavelength of the free surface motion is generally smaller than in geophysically generated tsunamis. Thus, dispersive effects need to be taken into account. In the present study the fluid layer is modeled by the Peregrine system modified appropriately and written in conservative variables. The landslide is assumed to be a quasi-deformable body of mass whose trajectory is completely determined by its barycenter motion. A differential equation modeling the landslide motion along a curvilinear bottom is obtained by projecting all the forces acting on the submerged body onto a local moving coordinate system. One of the main novelties of our approach consists in taking into account curvature effects of the sea bed.
\end{abstract}


\maketitle

\tableofcontents

\section{Introduction}

Extreme water waves can become an important hazard in coastal areas. Many geophysical mechanisms are related with underwater earthquakes and landslides. The former genesis mechanism has been intensively investigated since the Tsunami Boxing Day but also before \cite{Okal1988, Okal2003, Okal2004, Syno2006, Dutykh2006, Dutykh2007b, Beisel2009}. The list of references is far from being exhaustive. In this study we focus on the latter mechanism -- the underwater landslides which can cause some damage in the generation region. In general, the wavelength of landslide generated waves is much smaller than the length of transoceanic tsunamis. Consequently, the dispersive effects might be important. This consideration explains why we opt for a dispersive model \cite{Dutykh2010}, which is able to simulate the propagation and run-up of weakly nonlinear weakly dispersive water waves on nonuniform beaches.

Most of the landslide models which are currently used in the literature can be conventionally divided into three big categories. The first category contains the simplest models where the landslide shape and its trajectory are known {\em a priori} \cite{Tinti2001, todo2, MR2011541, Didenkulova2011}. Another approach consists in assuming that the landslide motion is translational and the sliding mass follows the trajectory of its barycenter. The governing equation of the center of mass is obtained by projecting all the forces, acting on the slide, onto the horizontal direction of motion \cite{Grilli1999, Watts2000, DiRisio2009}. Finally, the third category of models describe the slide-water evolution as a two-layer system, the sliding mass being generally described by a Savage-Hutter type model \cite{Fernandez-Nieto2007}. Taking into account all the uncertainties which exist in the modeling of the real-world events, we choose in this paper to study the intermediate level (i.e. the second category) which corresponds better to the precision of the available data.

The present study is organized as follows. In Section \ref{sec:water} we briefly describe the water wave model we use, while Section \ref{sec:slide} contains a more detailed presentation of the landslide model. The test-case considered in our study along with numerical results are presented in Section \ref{sec:num}. Finally, the main conclusions of this study are outlined in Section \ref{sec:concl}.

\section{Water wave model}\label{sec:water}

The water wave model we use in this study is based on the classical system derived by D.H.~Peregrine \cite{Peregrine1967}. However, the original derivation assumes that the bottom is stationary in time, i.e. $z=-d(x)$. Later, the bottom dynamics has been included into this system derivation by T.~Wu \cite{Wu1981,Wu1987}. In order to simulate the wave run-up, a conservative form of this system has to be derived. In the static bottom case it was done recently \cite{Dutykh2010}. The conservative system we use in the present study can be obtained in a similar way and can be written in the form:
\begin{equation}\label{eq:mper1}
  H_t + Q_x = 0,
\end{equation}
\begin{multline}\label{eq:mper2}
  \bigl(1+\frac13 H_x^2 - \frac16HH_{xx}\bigr)Q_t - \frac13H^2Q_{xxt} - \frac13 HH_xQ_{xt}
   + \Bigl(\frac{Q^2}{H} + \frac{g}{2}H^2\Bigr)_x = gH d_x + \frac12 H d_{xtt},
\end{multline}
where $H(x,t) := d(x,t) + \eta(x,t)$ is the total water depth and $\eta(x,t)$ is the free surface elevation below the still water level. The horizontal mass flux is denoted by $Q(x,t) := H(x,t)u(x,t)$ where $u(x,t)$ is the depth-averaged horizontal velocity variable. The bottom motion enters into the momentum balance equation (\ref{eq:mper2}) through the source term $\frac12 H d_{xtt}$. The mass conservation equation (\ref{eq:mper1}) keeps naturally its initial form. We underline that the linear dispersion relation of the modified Peregrine system (\ref{eq:mper1}), (\ref{eq:mper2}) is identical with that the original Peregrine model \cite{Peregrine1967} since these models differ only in nonlinear terms.

This modified Peregrine (m-Peregrine) system (\ref{eq:mper1}), (\ref{eq:mper2}) has several advantages. First of all, we note that the full water wave problem is invariant under the vertical translations \cite{Benjamin1982}. The asymptotic expansion method around the mean water level breaks this symmetry. The introduction of conservative variables $(H, Q)$ allows to recover this property. Another advantage is that the dispersive terms in the m-Peregrine system naturally vanish as the total water depth $H$ along with its first derivative $H_x$ tend to zero. This property is in complete agreement with the physical behavior of water waves which become more and more nonlinear to the detriment of the dispersion while approaching the shore.

In order to solve numerically the m-Peregrine system (\ref{eq:mper1}), (\ref{eq:mper2}) we choose the use of the finite volume method. Moreover, the run-up technique is well understood in the framework of Nonlinear Shallow Water equations \cite{Dutykh2009a}, which allows us to reuse this technology in the dispersive setting. The advective terms are discretized using the FVCF approach \cite{Ghidaglia2001} with UNO2 space reconstruction \cite{HaOs}. The dispersive terms are treated with the finite differences. For the time discretization we use the Bogacki-Shampine 3rd order Runge-Kutta scheme with adaptive time step control. Note that on each time step we have to solve a tridiagonal system of linear equations in order to determine the time derivative $Q_t$. We refer to \cite{Dutykh2010} for more details on the numerical method.

\section{Landslide model}\label{sec:slide}

In this section we briefly present a model of an underwater landslide motion. This process has to be addressed carefully since it determines the subsequent formation of water waves. In this study we will assume the moving mass to be a solid quasi-deformable body with a prescribed shape and known physical properties that preserves its mass and volume. Under these assumptions it is sufficient to compute the trajectory of the barycenter $x = x_c(t)$ to determine the motion of the whole body. In general, only uniform slopes are considered in the literature in conjunction with this type of landslide models \cite{Pelinovsky1996,Grilli1999,Watts2000,DiRisio2009}. However, a novel model, taking into account the bottom geometry and curvature effects, has been recently proposed \cite{KhakimzyanovG.S.Shokina2010}. Hereafter we will follow in great lines this study.

The static bathymetry is prescribed by a sufficiently smooth (at least of the class $C^2$) and single-valued function $z = -d_0(x)$. The landslide shape is initially prescribed by a localized in space function $z = \zeta_0(x)$. For example, in this study we choose the following shape function:
\begin{equation}\label{eq:cos}
  \zeta_0(x) = A\left\{
  \begin{array}{lc}
  \frac{1}{2}\Bigl(1+\cos(\frac{2\pi(x-x_0)}{\ell})\Bigr), & |x - x_0| \leq \frac{\ell}{2}\\
  0, & |x - x_0| > \frac{\ell}{2},
  \end{array}
  \right.
\end{equation}
where parameters $A$ is the maximum height, $\ell$ is the length of the slide and $x_0$ is the initial position of its barycenter. Obviously, the model description given below is valid for any other reasonable shape.

Since the landslide motion is translational, its shape at time $t$ is given by the function $z = \zeta(x,t) = \zeta_0(x-x_c(t))$. Recall that the landslide center is located at the point with abscissa $x = x_c(t)$. Then, the impermeable bottom for the water wave problem can be easily determined at any time by simply superposing the static and dynamic components:
\begin{equation*}
  z = -d(x,t) = -d_0(x) + \zeta(x,t).
\end{equation*} 

To simplify the subsequent presentation, we introduce the classical arc-length parametrization, where the parameter $s = s(x)$ is given by the following formula:
\begin{equation}\label{eq:len}
  s = L(x) = \int_{x_0}^{x}\sqrt{1+(d_0'(\xi))^2}\,d\xi.
\end{equation}
The function $L(x)$ is monotonic and can be efficiently inverted to turn back to the original Cartesian abscissa $x = L^{-1}(s)$. Within this parametrization, the landslide is initially located at point with the curvilinear coordinate $s = 0$. The local tangential direction is denoted by $\tau$ and the normal by $n$.

The landslide motion is governed by the following differential equation obtained by a straightforward application of Newton's second law:
\[
  m\od{^2s}{t^2} = F_\tau(t),
\]
where $m$ is the mass and $F_\tau(t)$ is the tangential component of the forces acting on the moving submerged body. In order to project the forces onto the axes of local coordinate system, the angle $\theta(x)$ between $\tau$ and $Ox$ can be easily determined:
\[
  \theta(x) = \arctan\bigl(d'_0(x)\bigr).
\]

Let us denote by $\rho_w$ and $\rho_\ell$ the densities of the water and sliding material correspondingly. If $V$ is the volume of the slide, then the total mass $m$ is given by this expression:
\[
  m := (\rho_\ell + c_w\rho_w)V,
\]
where $c_w$ is the added mass coefficient \cite{Batchelor2000}. A portion of the water mass has to be added since it is entrained by the underwater body motion. For a cylinder, for example, the coefficient $c_w$ is equal exactly to one. The volume $V$ can be computed as
\[
  V = W\cdot S = W\int_{\R}\zeta_0(x) \, dx,
\]
where $W$ is the landslide width in the transverse direction. The last integral can be computed exactly for the particular choice (\ref{eq:cos}) of the landslide shape to give
\[
  V = \frac12\ell A W.
\]

The total projected force $F_\tau$ acting on the landslide can be conventionally represented as a sum of two different kind of forces denoted by $F_g$ and $Fd$:
\[
  F_\tau = F_g + F_d,
\]
where $F_g$ is the joint action of the gravity and buoyancy, while $F_d$ is the total contribution of various dissipative forces (to be specified below).

The gravity and buoyancy forces act in opposite directions and their horizontal projection $F_g$ can be easily computed:
\[
  F_g(t) = (\rho_\ell - \rho_w)W g\int_\R \zeta(x,t)\sin\bigl(\theta(x)\bigr)\,dx.
\]
Now, let us specify the dissipative forces. The water resistance to the motion force $F_r$ is proportional to the maximal transversal section of the moving body and to the square of its velocity:
\[
  F_r = -\frac12 c_d\rho_w AW\sigma(t)\Bigl(\od{s}{t}\Bigr)^2,
\]
here $c_d$ is the resistance coefficient of the water and $\sigma(t) := \sign\Bigl(\od{s}{t}\Bigr)$. The coefficient $\sigma(t)$ is needed to dissipate the landslide kinetic energy independently of its direction of motion. The friction force $F_f$ is proportional to the normal force exerted on the body due to the weight:
\[
  F_f = -c_f \sigma(t)N(x,t).
\]
The normal force $N(x,t)$ is composed of the normal components of gravity and buoyancy forces but also of the centrifugal force due to the variation of the bottom slope:
\begin{equation*}
  N(x,t) = (\rho_\ell - \rho_w)g W\int_\R\zeta(x,t)\cos\bigl(\theta(x)\bigr)\,dx + \rho_\ell W\int_\R\zeta(x,t)\kappa(x)\Bigl(\od{s}{t}\Bigr)^2\, dx,
\end{equation*}
where $\kappa(x)$ is the signed curvature of the bottom which can be computed by the following formula:
\begin{equation*}
  \kappa(x) = \frac{d''_0(x)}{\bigl(1+(d'_0(x))^2\bigr)^{\frac32}}.
\end{equation*}
We note that the last term vanishes for a plane bottom since $\kappa(x) \equiv 0$ in this particular case.

Finally, if we sum up all the contributions of described above forces, we obtain the following second order differential equation:
\begin{equation}\label{eq:ODE}
  (\gamma + c_w)S\od{^2s}{t^2} = (\gamma-1)g \Bigl(I_1(t)-c_f\sigma(t)I_2(t)\Bigr) - \sigma(t)\Bigl(c_f\gamma I_3(t) + \frac12 c_d A\Bigr)\Bigl(\od{s}{t}\Bigr)^2,
\end{equation}
where $\gamma := \frac{\rho_\ell}{\rho_w} > 1$ is the ratio of densities and integrals $I_{1,2,3}(t)$ are defined as:
\[
  I_1(t) = \int_\R\zeta(x,t)\sin\bigl(\theta(x)\bigr)\,dx,
\]
\[
  I_2(t) = \int_\R\zeta(x,t)\cos\bigl(\theta(x)\bigr)\,dx,
\]
\[
  I_3(t) = \int_\R \zeta(x,t)\kappa(x)\,dx.
\]
Note also that equation (\ref{eq:ODE}) was simplified by dividing both sides by the width value $W$. In order to obtain a well-posed initial value problem, equation (\ref{eq:ODE}) has to be completed by two initial conditions:
\[
  s(0) = 0, \quad s'(0) = 0.
\]
In order to solve numerically equation (\ref{eq:ODE}) we employ the same Bogacki-Shampine 3rd order Runge-Kutta scheme as we use to solve the Boussinesq equations (\ref{eq:mper1}), (\ref{eq:mper2}). The integrals $I_{1,2,3}(t)$ are computed using the trapezoidal rule. Once the landslide trajectory $s = s(t)$ is found, we use equation (\ref{eq:len}) to find its motion $x = x(t)$ in the initial Cartesian coordinate system.

\section{Numerical results}\label{sec:num}

\begin{figure}
  \centering
  \includegraphics[width=0.82\textwidth]{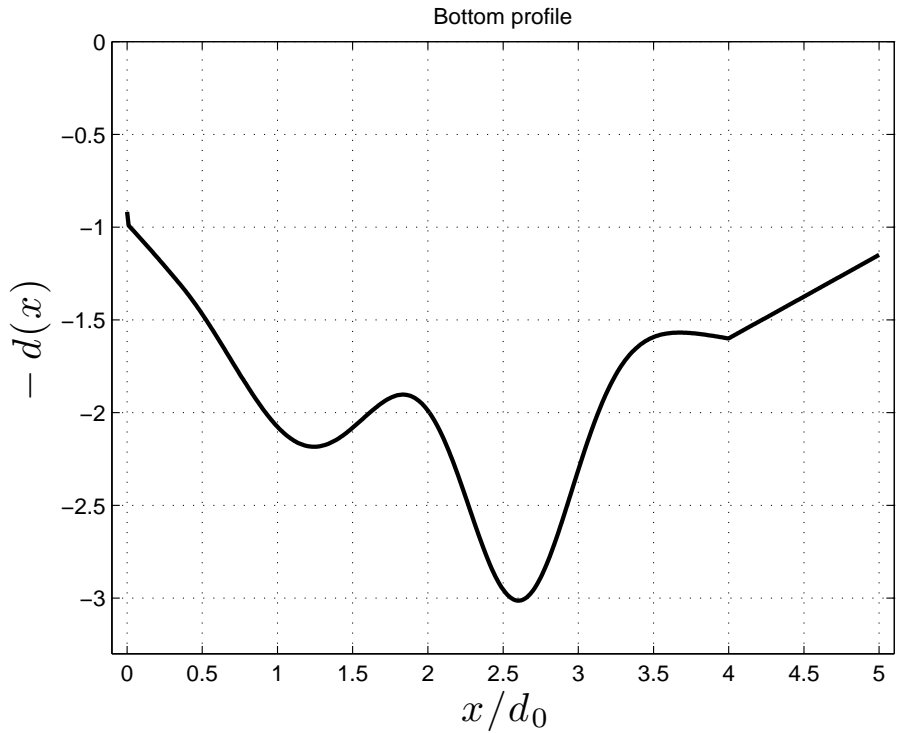}
  \caption{Bathymetry profile for parameters given in Table \ref{tab:pars}.}
  \label{fig:bot}
\end{figure}

Let us consider the one-dimensional computational domain $\I = [a, b] = [-3, 10]$ composed from three regions: the left and right sloping beaches surrounding a complex generation region. Specifically, the static bathymetry function $d_0(x)$ is given by the following expression:
\[
  d_0(x) = \left\{
  \begin{array}{ll}
    d_0 + 4\tan\delta\cdot x, & x \leq 0, \\
    d_0 + \tan\delta\cdot x + p(x), & 0 < x \leq m, \\
    d_0 + 4m\tan\delta - 3\tan\delta\cdot x, & x > m, \\
  \end{array}
  \right.
\]
where the function $p(x)$ is defined as
\[ 
  p(x) =  A_1e^{-k_1(x-x_1)^2} + A_2e^{-k_2(x-x_2)^2}.
\]
Basically, this function represents a perturbation of the sloping bottom by two underwater bumps. We made this nontrivial choice in order to illustrate better the advantages of our landslide model, which was designed to handle general non-flat bathymetries. The values of all physical and numerical parameters are given in Table \ref{tab:pars}. The bottom profile for these parameters is depicted on Figure \ref{fig:bot}.

\begin{table}
 \centering
 \begin{tabular}{c|c}
   Parameter & Value \\
   \hline\hline
   Gravity acceleration, $g$ & $1.0$ \\
   Water depth at $x=0$, $d_0$ & $1.0$ \\
   Bottom slope, $\tan(\delta)$ & $0.15$ \\
   Underwater bump amplitude, $A_1$ & $1.0$ \\
   Underwater bump amplitude, $A_2$ & $1.6$ \\
   Bump steepness, $k_1$ & $1.9$ \\
   Bump steepness, $k_2$ & $3.9$ \\
   Bump center position, $x_1$ & $1.2$ \\
   Bump center position, $x_2$ & $2.6$ \\
   Boundary between bottom regions, $m$ & $4.0$ \\
   Number of control volumes, $N$ & $1000$ \\
   Slide amplitude, $A$ & $0.3$ \\
   Length of the slide, $\ell$ & $2.0$ \\
   Initial slide position, $x_0$ & $0.0$ \\
   Added mass coefficient, $c_w$ & $1.0$ \\
   Water drag coefficient, $c_d$ & $1.0$ \\
   Friction coefficient, $c_f$ & $\tan1^\circ$ \\
   Ratio between water and slide densities, $\gamma$ & 1.5 \\
   \hline\hline
 \end{tabular}
 \caption{Values of various parameters used in the numerical computations.}
 \label{tab:pars}
\end{table}

\begin{figure}
  \centering
  \includegraphics[width=0.82\textwidth]{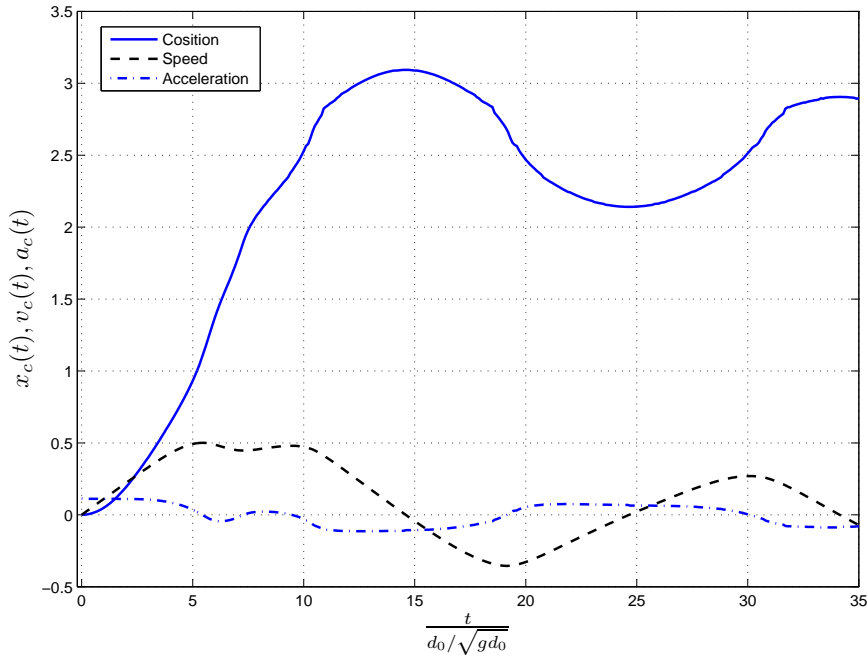}
  \caption{Barycenter position (blue solid line), velocity (black dashed line) and acceleration (blue dash-dotted line) during the landslide motion.}
  \label{fig:slide}
\end{figure}

The landslide motion starts from the rest position under the action of the gravity force. We study its motion along with the waves of the free surface up to $T=21$ s. The landslide barycenter trajectory along with its speed and acceleration are shown in Figure \ref{fig:slide}. As it is expected, the landslide remains trapped in the second underwater bump, where it oscillates before stopping completely its motion.

\begin{figure}
  \centering
  \includegraphics[width=0.82\textwidth]{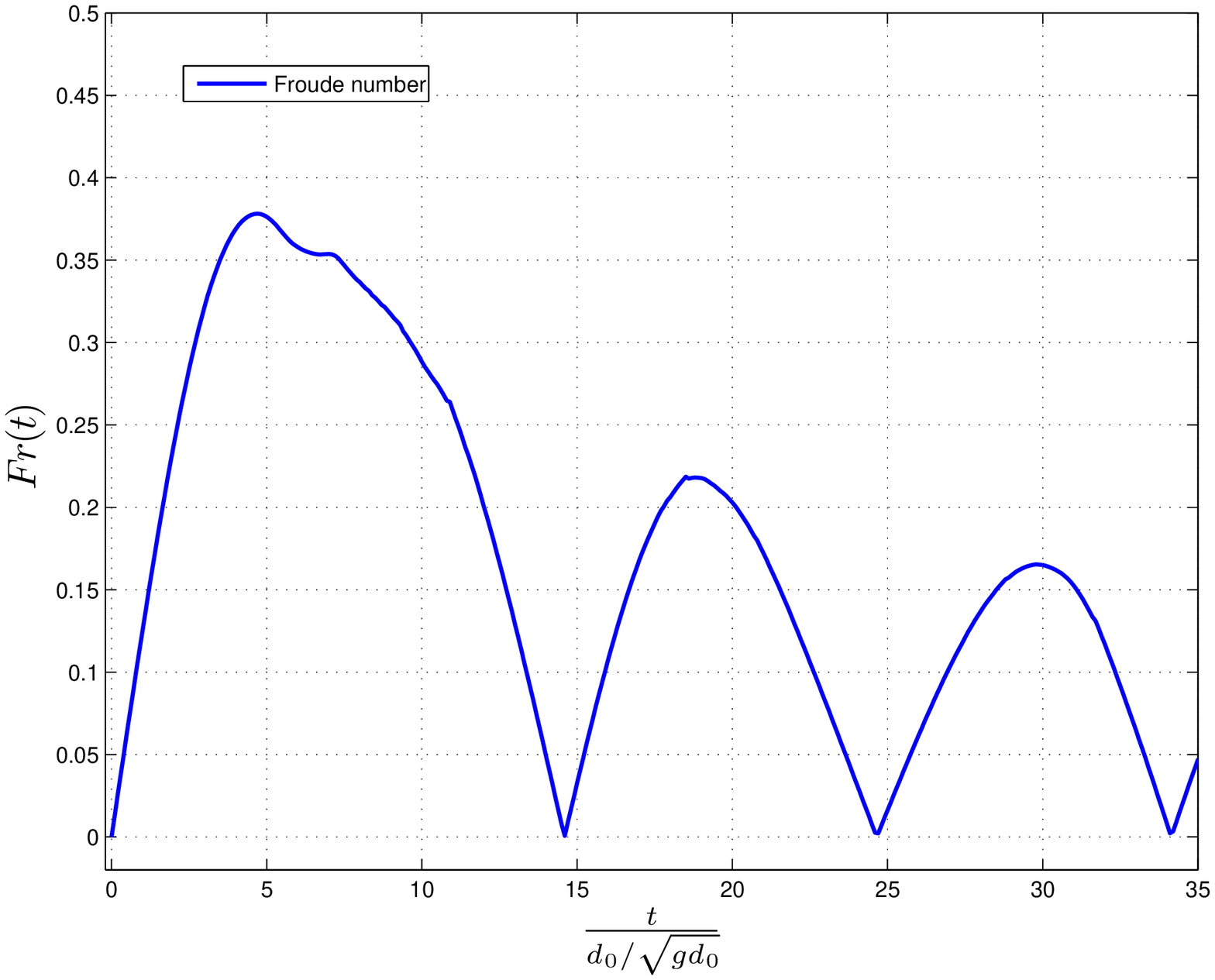}
  \caption{Local Froude number computed along the slide motion.}
  \label{fig:froude}
\end{figure}

One of the important parameters in shallow water flows is the Froude number defined as the ratio between the characteristic fluid velocity to the gravity wave speed. We computed also this parameter along the landslide trajectory:
\[
  \mathrm{Fr}(t) := \frac{|x'_c(t)|}{\sqrt{gd\bigl(x_c(t), t)\bigr)}}.
\]
The result is presented in Figure \ref{fig:froude}. We can see that in our case the motion remains subcritical during the experiment.

\begin{figure}
  \centering
  \includegraphics[width=0.82\textwidth]{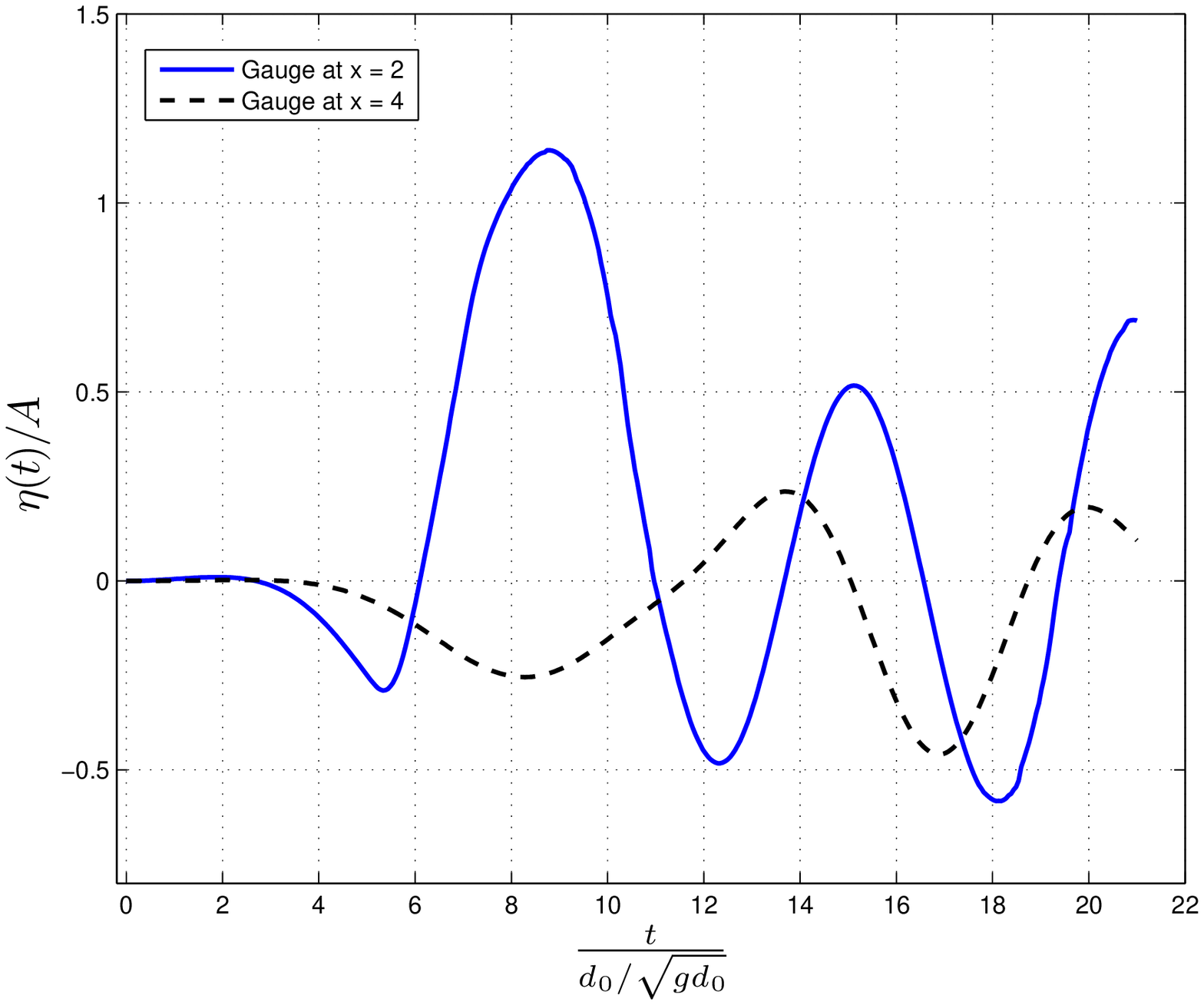}
  \label{fig:gauge}
  \caption{Synthetic wave gauge records located at $x = 2$ (blue solid line) and $x = 4$ (black dashed line). The vertical axis is relative to the landslide amplitude $A$.}
\end{figure}

\begin{figure}
  \centering
  \includegraphics[width=0.82\textwidth]{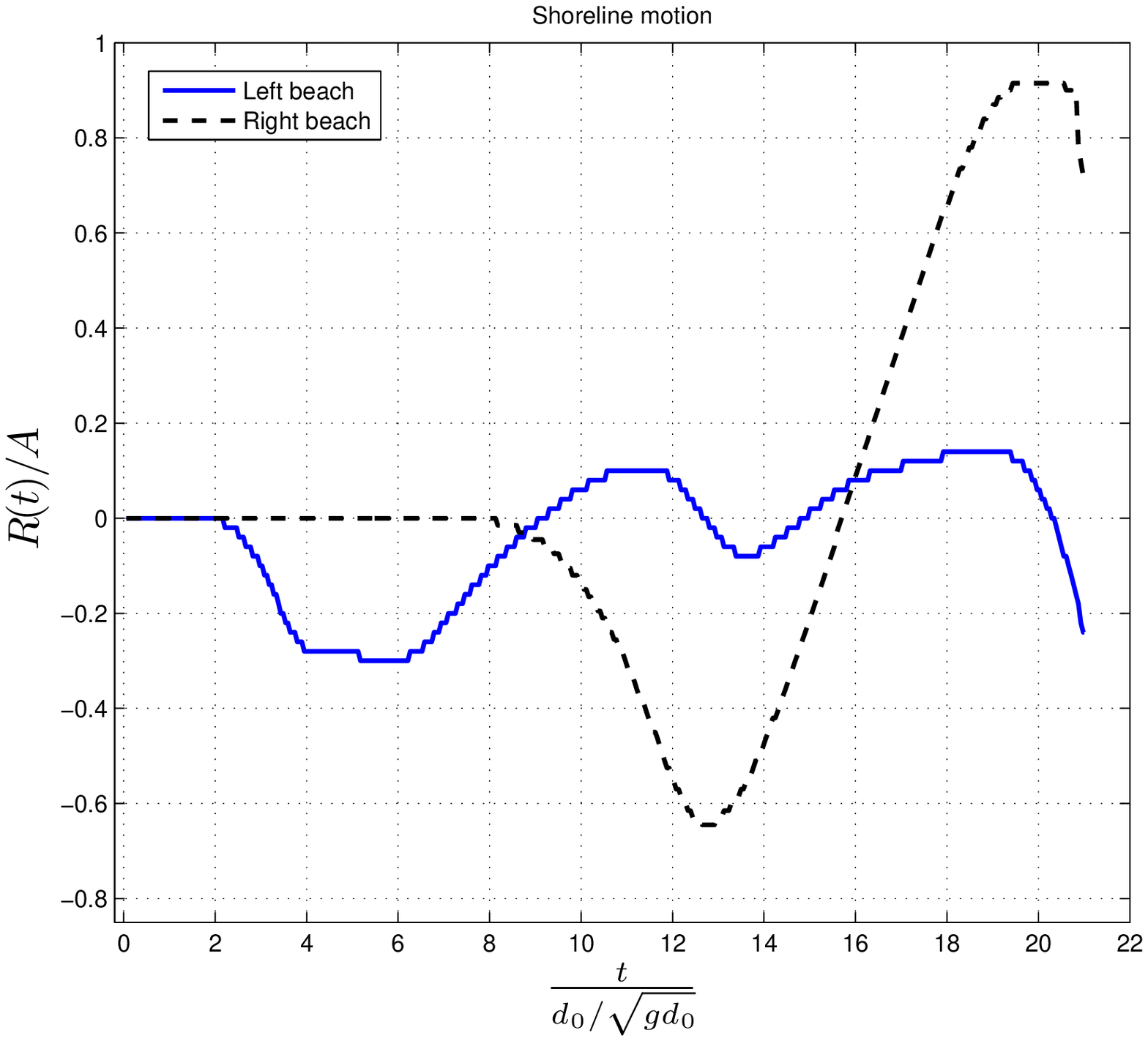}
  \caption{Shoreline motion on the left (blue solid line) and right (black dashed line) beaches.}
  \label{fig:rup}
\end{figure}

We installed two synthetic wave gauges located at $x = 2$ (point located between two underwater bumps) and $x = m$ (endpoint of the generation region and the beginning of the right sloping beach, $m = 4$ in our simulations) in order to measure the magnitude of generated free surface motions. These synthetic records are presented in Figure \ref{fig:gauge}. Finally, the wave run-up is simulated numerically on the left and right beaches. The shoreline motion is represented in Figure \ref{fig:rup}. We can see, that the proposed scenario provides higher run-up values on the opposite beach to the slope where the landslide takes place. 

\section{Conclusions}\label{sec:concl}

In this study we presented a novel model of a landslide motion over general curvilinear bottoms. This model takes into account the effects of bottom curvature, generally neglected in the literature \cite{Pelinovsky1996,Grilli1999,Watts2000,DiRisio2009}. Despite the inclusion of some new physical effects, the considered model is computationally inexpensive and can be potentially used in more operational context. The computed bottom motion is strongly coupled with a conservative Peregrine system \cite{Dutykh2010} which describes the propagation of nonlinear weakly dispersive water waves. The run-up of landslide generated waves on both beaches is simulated using the m-Peregrine system. The extension of this approach to three dimensions represents a natural perspective for future research along with investigation of other possible scenarios.

\section*{Acknowledgements}

S.~Beisel, N.~Shokina and D.~Dutykh acknowledge the support from the projects PICS CNRS No 5607 and RFBR No 10-05-91052. D.~Dutykh also acknowledges the French Agence Nationale de la Recherche, project MathOc\'ean (ANR-08-BLAN-0301-01). The authors would like to thank Professors Leonid Chubarov and Gayaz Khakimzyanov for very helpful and stimulating discussions on the landslide modeling.

%%% Bibliography %%%
\bibliography{mybib}
\bibliographystyle{alpha}

\end{document}